\documentclass[draftclsnofoot,12pt, onecolumn]{IEEEtran}
\usepackage{amsmath}
\usepackage{epsfig}
\usepackage{amssymb}
\usepackage{latexsym}
\usepackage{color}
\usepackage{graphicx}
\begin{document}
\title{Tight Approximations for the  Two Dimensional Gaussian $Q-$function}
\author{\IEEEauthorblockN{Paschalis C. Sofotasios\\}
\IEEEauthorblockA{School of Electronic and Electrical Engineering \\
University of Leeds, UK\\
Email: eenpso@leeds.ac.uk\\}
\and
\IEEEauthorblockN{Steven Freear\\}
\IEEEauthorblockA{School of Electronic and Electrical Engineering\\
University of Leeds, UK\\
Email: s.freear@leeds.ac.uk}}
\maketitle 
\begin{abstract} 
The aim of this work is the derivation of two approximated expressions for the two dimensional Gaussian Q-function, $Q(x,y;\rho)$. These expressions are highly accurate and are expressed in closed-form. Furthermore, their algebraic representation is relatively simple and therefore, convenient to handle both analytically and numerically. This feature is particularly useful for two reasons: firstly because it renders the derived expressions useful mathematical tools that can be utilized in numerous analytic performance evaluation studies in digital communications under fading; secondly because the two dimensional Gaussian Q-function is neither tabulated nor a built-in function in popular mathematical software packages such as $Maple$, $Mathematica$ and $Matlab$. \end{abstract}
\section{Introduction}
$ $\\
\indent
Special functions are invaluable mathematical tools in natural sciences and engineering. Their wide use in analytical performance evaluation studies in wireless communications often allows the derivation of closed-form relationships for vital performance measures such as channel capacity and probability of error. In addition, the majority of special functions are built-in functions in popular mathematical software packages. As such, the algebraic representation of any related expressions as well as their computation have been significantly simplified. \\
\indent
It is recalled that special functions play a significant role in   the area of digital communications [11]-[13] and [16]-[19] and the references therein. Such a special function is also the two dimensional Gaussian Q-function. It is denoted as $Q(x,y;\rho)$ and is constituted by the two-dimensional Gaussian integral, [1]-[5]. Alternative integral representations for $Q(x,y;\rho)$ along with performance bounds were reported in [6]-[10]. Furthermore, an exact infinite series representation, which is expressed in terms of the incomplete gamma function and elementary functions, was derived in [11]-[12]. Nevertheless, it is noted that this series is the only explicit relationship for the two dimensional Gaussian Q-function since $Q(x,y;\rho)$ is neither expressed in terms of other elementary and/or special functions, nor it is included as a built-in function in popular mathematical software packages such as $Maple$, $Mathematica$ and $Matlab$. As a consequence, both its computation and analytical tractability are not as straightforward as in the case of most special functions in communication theory. \\
\indent
Motivated by this, the aim of this work is the derivation of two tight approximations for the $Q(x,y;\rho)$ function. The high accuracy of the offered expressions is validated through comparisons with results obtained from numerical integrations as well as by the identity $Q(x,y;\rho) \triangleq Q(x)Q(y)$. An important feature of the proposed approximations is that they are expressed in closed-form and have a rather simple algebraic representation. As a result, they are expected to be useful in various analytical studies relating to the performance evaluation of digital communications over fading environments.

\section{The Two dimensional Gaussian Q-function}

\indent
The two dimensional Gaussian Q-function is defined as: 

\begin{equation} 
Q(x,y; \rho) \triangleq \frac{1}{2 \pi \sqrt{1 - \rho ^{2}}} \int_{x}^{\infty} \int_{y} ^{\infty} e^{-\frac{u^{2} + v^{2} - 2 \rho u v}{2 \left(1 - \rho ^{2} \right)}}du dv
\end{equation}
According to [2], the above relationship can be also represented in the more desired Craig form as follows,
 
\begin{equation} 
Q(x, y; \rho) = \frac{1}{2 \pi} \int_{0}^{tan^{-1} \left(\frac{\sqrt{1 - \rho^{2}} x/y}{1 - \rho x/y} \right)} e^{- \frac{x^{2}}{2 sin^{2}\Phi}} d \Phi  + \, \frac{1}{2 \pi} \int_{0}^{tan^{-1} \left(\frac{\sqrt{1 - \rho^{2}} y/x}{1 - \rho y/x} \right)} e^{- \frac{y^{2}}{2 sin^{2}\Phi}} d \Phi 
\end{equation}
For the special case that $x = y$, equation $(2)$ reduces to [2],

\begin{equation} 
Q(x, x; \rho) = \frac{1}{\pi} \int_{0}^{tan^{-1} \left(\sqrt{\frac{1 + \rho}{1 - \rho} } \right)} e^{- \frac{x^{2}}{2 sin^{2} \phi}} d \Phi
\end{equation}
whereas for the case that $\rho = 0$, it reduces to the product of the one-dimensional Gaussian Q-functions of its arguments, namely, 

\begin{equation} 
Q(x, y; \rho = 0) = Q(x)Q(y)
\end{equation}
An exact series representation for $Q(x,y;\rho)$ was reported in [11]-[12], namely,
 
\begin{equation} 
Q(x, y ;\rho) = \frac{Q(x)}{2} -  \frac{1}{\pi} \sum_{l = 0}^{\infty} \sum_{k = 0}^{2l+1} \frac{(-1)^{3l+1-k} (2l)! \, y^{k}\, \rho^{2l + 1 - k} \,\Gamma\left( 1 + l - \frac{k}{2}, \frac{x^{2}}{2} \right)}{l!\, k! \left( 1 - \rho^{2}\right)^{l + 1/2}\, (2l + 1 - k)!\, 2^{1 + k/2}}
\end{equation} 

\section{Closed-form approximations for $Q(x,y;\rho)$}

\indent
It was mentioned above that the only explicit representation of $Q(x,y;\rho)$ is the infinite series in $(5)$. The algebraic form of this series is relatively simple. However, the fact that it has an infinite form typically raises convergence issues. More specifically, the number of terms required for truncating the series adequately is sufficiently related to its parameters. This is evident by the fact that for small values of $x$, $y$ or $\rho$, only a few number of terms are required to truncate the series with small error. On the contrary, as as the value of $y$ and $\rho$ increases, the numbers of terms required for a small truncation error also increase. As a consequence, the required computation time increase as well.\\
\indent
An effective mean to resolve such issues is the derivation accurate approximations that are expressed in closed-form and are also analytically tractable. Such approximated expression can be ultimately derived with the aid of two approximations for the one dimensional Q-function, $Q(x)$, that were proposed in [11]-[13], namely,

\begin{equation} 
Q(x) \simeq 0.49e^{-\frac{8}{13}x} e^{-\frac{x^{2}}{2}}
\end{equation}
and

\begin{equation} 
Q(x) \simeq 0.208 e^{-0.876 x^{2}} + 0.13 e^{- 0.525x^{2}} + 0.14 e^{-7.25 x^{2}}
\end{equation}
\begin{figure}[h]
\centerline{\psfig{figure=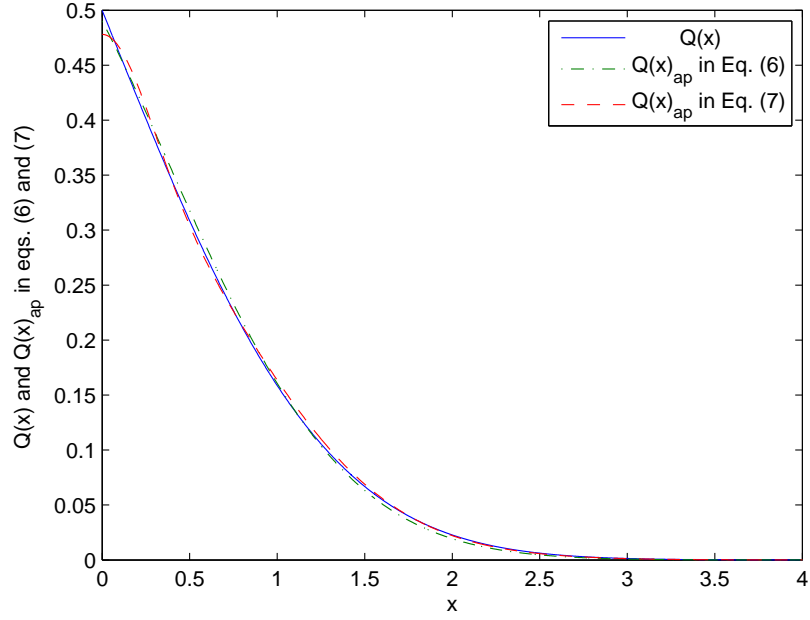,width=12cm,height=9cm}}
\caption{Comparison of $\hat{Q}(x)$ in eqs. $(6)$ and $(7)$ with $Q(x)$ in $Matlab$}
\end{figure}
The behaviour and high accuracy of $(6)$ and $(7)$ with respect to $Q(x)$ are illustrated in figure $1$. This can be also justified by figure $2$ which depicts the level of the involved absolute error and absolute relative error, namely, 

\begin{equation} 
\epsilon_{a} = \mid Q(x) - \hat{Q}(x) \mid
\end{equation}
and

\begin{equation} 
\epsilon_{ar} = \frac{\mid Q(x) - \hat{Q}(x) \mid}{Q(x)}
\end{equation}
\begin{figure}[h]
\centerline{\psfig{figure=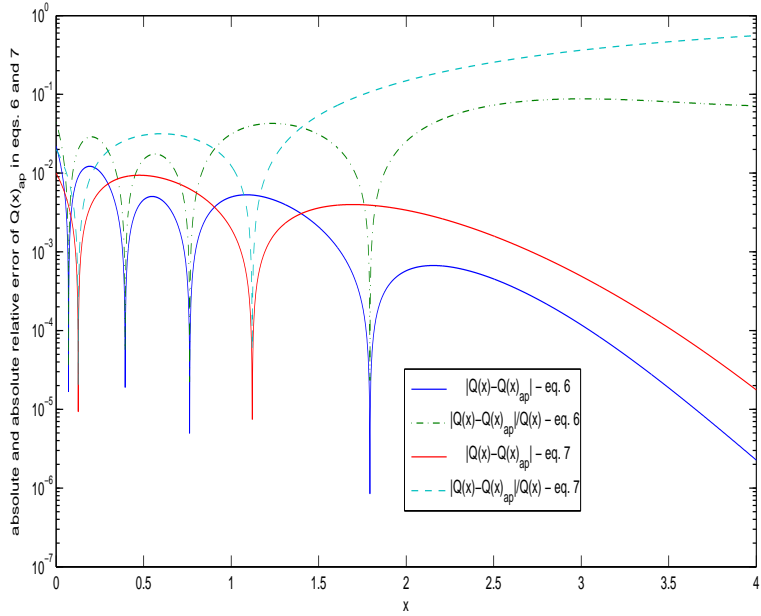,width=12cm,height=9cm}}
\caption{Absolute error, $\epsilon_{a}$, and absolute relative error, $\epsilon_{ar}$, between $Q(x)$ and eqs. $(6)$ and $(7)$}
\end{figure}
 
\subsection{A first approximation for $Q(x,y;\rho)$ function}
 
A closed-form approximation for the $Q(x,y;\rho)$ function can be derived as follows: The double integral in $(1)$ can be alternatively expressed as, [11],
 
\begin{equation} 
Q(x, y; \rho) = \frac{1}{2\pi \sqrt{1 - p^{2}}}   \int_{x}^{\infty} e^{- \frac{v^{2}}{2 \left(1 - \rho^{2} \right)}} \left[\int_{y}^{\infty} e^{- \frac{u^{2}}{2 \left(1 - \rho^{2} \right)}} e^{\frac{\rho uv}{\left(1 - \rho^{2} \right)}} du\right]dv 
\end{equation}
Consequently, by integrating once by part the inner integral and carrying out some straightforward algebraic manipulations, the following equivalent representation is deduced,

\begin{equation} 
Q(x, y; \rho) = \frac{1}{\sqrt{2 \pi}} \int_{x}^{\infty} e^{-\frac{v^{2}}{2}} Q\left( \frac{y - \rho v}{\sqrt{1 - \rho^{2}}} \right) dv
\end{equation}
Thus, it is evident that an explicit expression for $Q(x, y; \rho)$ is subject to analytic evaluation of the semi-infinite integral,

\begin{equation} 
\int_{a}^{\infty} e^{-b x^{2}} Q(cx - d)dx
\end{equation}
where $ a, b, c, d \in \mathbb{R}$. To this effect, by substituting $(6)$ into $(11)$, one obtains

\begin{equation} 
Q(x, y; \rho) \simeq \frac{0.49}{\sqrt{2 \pi}} \int_{x}^{\infty} e^{-\frac{v^{2}}{2}} e^{-\frac{8}{13} \frac{y - \rho v}{\sqrt{1 - \rho ^{2}}}} e^{-\frac{(y - \rho v)^{2}}{2(1 - \rho ^{2})}} dv
\end{equation}
which after algebraic manipulations yields the following closed-form expression, 
\begin{figure}[h]
\centerline{\psfig{figure=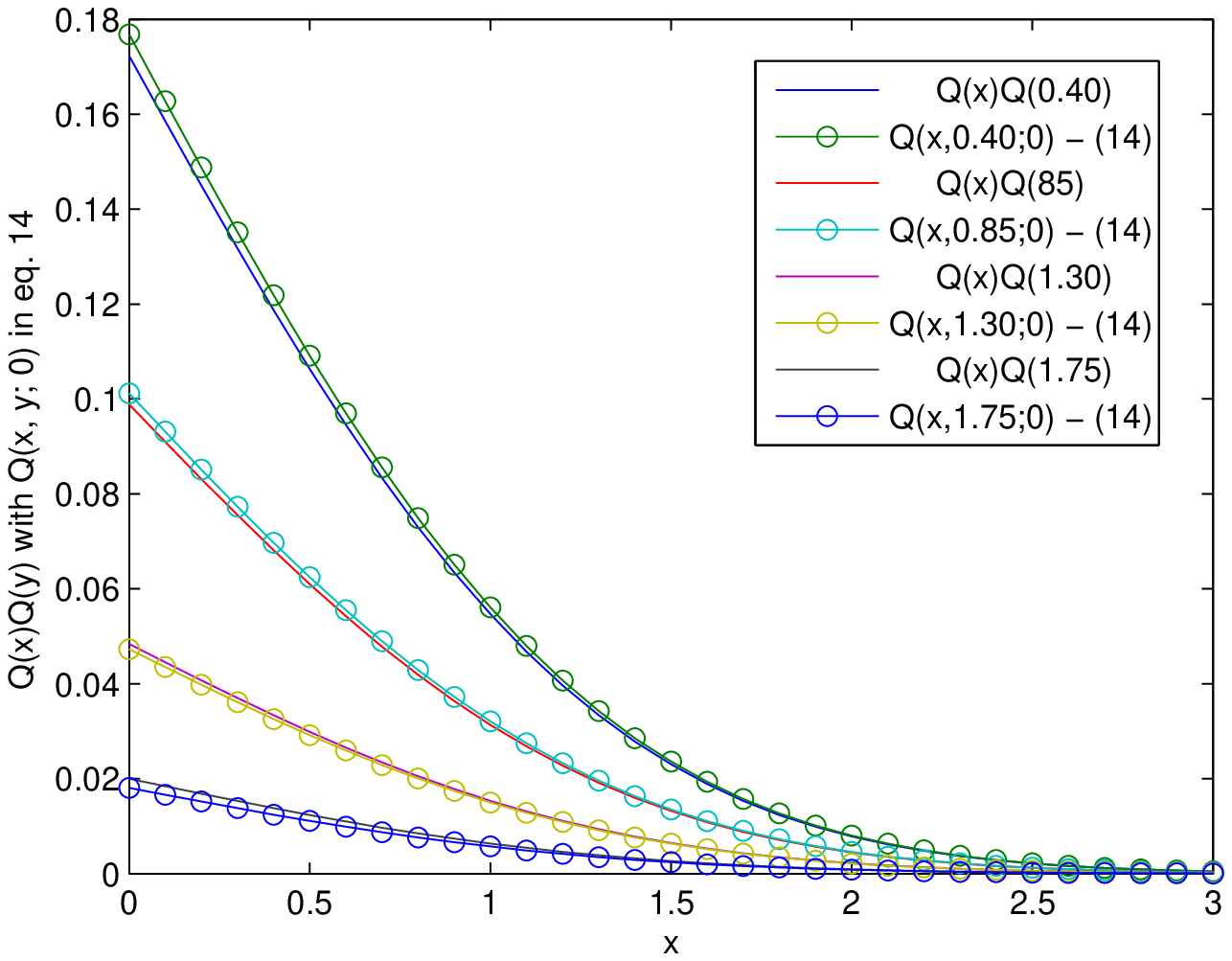,width=12cm,height=9cm}}
\caption{Comparison of $Q(x)Q(y)$ with $Q(x, y;\rho = 0)$ in eq. $14$ }
\end{figure}

\begin{equation} 
Q(x, y, \rho) \simeq \frac{0.49}{\sqrt{A}} e^{-\frac{8y}{13\sqrt{1-\rho^{2}}}} e^{-\frac{y^{2}}{2 \left(1 - \rho^{2} \right)}} e^{\frac{B^{2}}{2A}} Q \left(x\sqrt{A} - \frac{B}{\sqrt{A}} \right)
\end{equation}
where,

\begin{equation}
A = 1 + \frac{\rho^{2}}{1 - \rho^{2}} 
\end{equation}
and

\begin{equation}
 B = \frac{8\rho}{13\sqrt{1 - \rho^{2}}} + \frac{\rho y}{1 - \rho ^{2}} 
\end{equation}

\begin{figure}[h]
\centerline{\psfig{figure=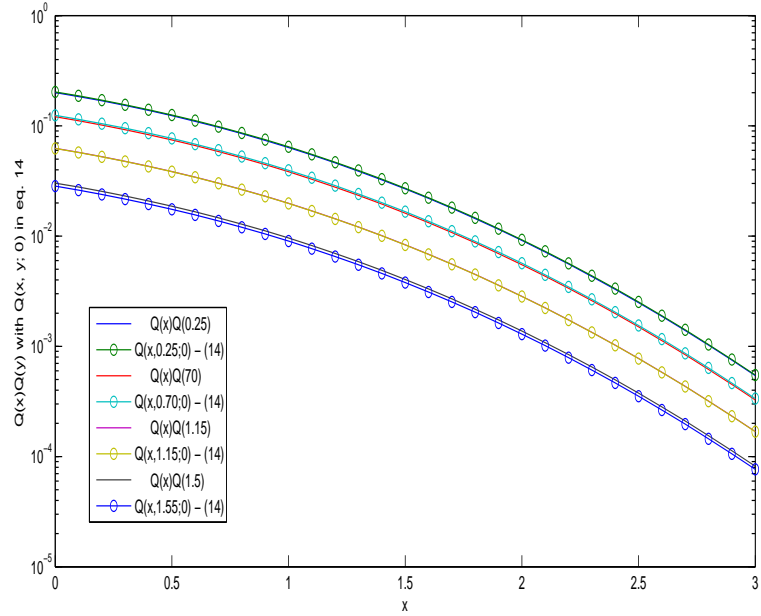,width=12cm,height=9cm}}
\caption{Comparison of $Q(x)Q(y)$ with $Q(x, y;\rho = 0)$ in eq. $14$ }
\end{figure}
It is recalled here that the $Q(x,y;\rho)$ function is not a built-in function in mathematical software packages such as $Maple$, $Mathematica$ and $Matlab$. However, the following identity holds by definition, $Q(x,y;\rho = 0) = Q(x)Q(y)$. As a consequence, the accuracy of equation $(14)$ is assessed by means of the this identity in figures $3$ and $4$. One can clearly observe the excellent agreement between the plotted curves for almost all cases. This is also evident by the level of the involved absolute and absolute relative which for all cases is less than $5 \%$.  
 
\subsection{A second approximation for $Q(x,y;\rho)$ function}
 
By following the same methodology, a second closed-form approximation for the two dimensional Q-function can be derived. To this end, by making the necessary transformation of variables in $(7)$ and then substituting in $(11)$, the following relationship is deduced, 

$$
Q(x, y; \rho) \simeq \frac{0.208}{\sqrt{2 \pi}} \int_{x}^{\infty} e^{-\frac{v^{2}}{2}} e^{-\frac{0.876}{1 - \rho^{2}} (y - \rho v)^{2}} dv$$
$$
\, \qquad \, \quad \, + \frac{0.13}{\sqrt{2 \pi}} \int_{x}^{\infty} e^{-\frac{v^{2}}{2}} e^{-\frac{0.525}{1 - \rho^{2}} (y - \rho v)^{2}} dv
$$
\begin{equation} 
\, \qquad \, \quad \, + \frac{0.14}{\sqrt{2 \pi}} \int_{x}^{\infty} e^{-\frac{v^{2}}{2}} e^{-\frac{7.25}{1 - \rho^{2}} (y - \rho v)^{2}} dv
\end{equation}
Evidently, the derivation of a closed-form expression for $(15)$ is subject to analytic evaluation of the three involved integrals. Notably, these integrals have the same algebraic form as the $(13)$. Therefore, after some straightforward algebraic manipulation, one obtains the following closed-form expression,

\begin{figure}[h]
\centerline{\psfig{figure=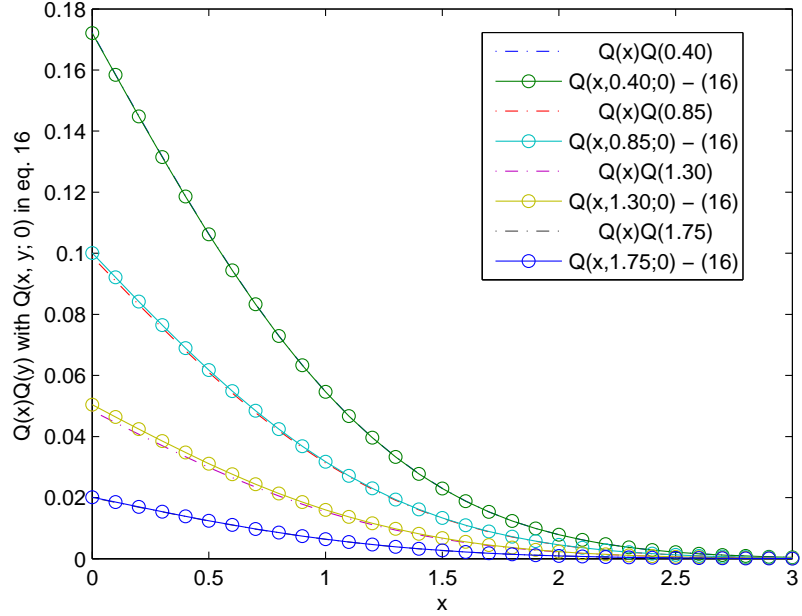,width=12cm,height=9cm}}
\caption{Comparison of $Q(x)Q(y)$ with $Q(x, y;\rho = 0)$ in eq. $16$ }
\end{figure}
The behaviour of $(16)$ is illustrated in figures $5$ and $6$ along with the $Q(x)$ function. One can observe the excellent agreements between the theoretical and approximated results. This agreement is also evident by the level of the involved absolute error and absolute relative error which is less than $4 \%$ for almost all parametric values. 
\begin{figure}[h]
\centerline{\psfig{figure=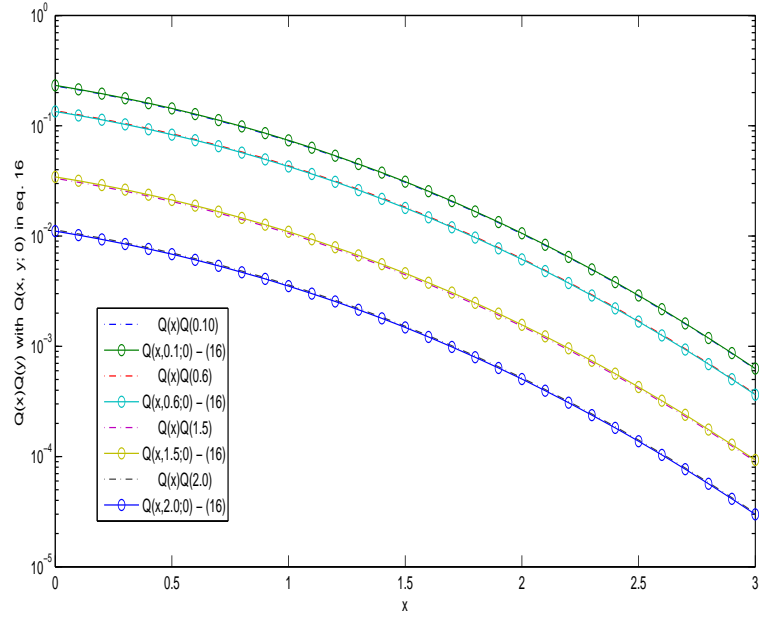,width=12cm,height=9cm}}
\caption{Comparison of $Q(x)Q(y)$ with $Q(x, y;\rho = 0)$ in eq. $16$ }
\end{figure}

$$
Q(x, y; \rho) \cong \frac{0.208}{\sqrt{2a}} e^{-\frac{0.876\,y^{2}}{1 - \rho^{2}}} e^{\frac{b^{2}\,y^{2}}{4a}} Q \left(x \sqrt{2a} - \frac{b\,y}{\sqrt{2a}} \right) 
$$
$$ + \frac{0.13}{\sqrt{2c}} e^{-\frac{0.525\,y^{2}}{1 - \rho^{2}}} e^{\frac{d^{2}\,y^{2}}{4c}} Q \left(x \sqrt{2c} - \frac{d\,y}{\sqrt{2c}} \right) $$
\begin{equation} 
+ \frac{0.14}{\sqrt{2f}} e^{-\frac{7.25\,y^{2}}{1 - \rho^{2}}} e^{\frac{g^{2}\,y^{2}}{4f}} Q \left(x \sqrt{2f} - \frac{g\,y}{\sqrt{2f}} \right) 
\end{equation}
where, 

\begin{equation}
a = \frac{1}{2} + \frac{0.876 \rho^{2}}{1 - \rho^{2}},
\end{equation}

\begin{equation}
b = \frac{1.752\rho}{1 - \rho^{2}},
\end{equation}

\begin{equation}
c = \frac{1}{2} + \frac{0.525 \rho^{2}}{1 - \rho^{2}},
\end{equation}

\begin{equation}
d = \frac{1.05\rho}{1 - \rho^{2}},
\end{equation}

\begin{equation}
f = \frac{1}{2} + \frac{7.25 \rho^{2}}{1 - p^{2}},
\end{equation}
and

\begin{equation}
g = \frac{14.5\rho}{1 -\rho^{2}}
\end{equation}

\section{Closing Remarks}
 
\indent
In this work, we derived two novel closed-form approximations for the two dimensional Gaussian Q-function, $Q(x,y;\rho)$. Two critical features of these expressions are their high accuracy and their relatively simple algebraic form. These features ultimately render them rather convenient to handle both analytically and numerically. The latter is particularly important since the two dimensional Gaussian Q-function is not a built-in function in popular mathematical software packages. As a consequence, the derived expression can be considered a useful mathematical tool in various analytical studies in natural sciences and engineering, in general, and particularly in applications in communication theory relating to digital communications over fading channels.  
 
\bibliographystyle{IEEEtran}

\end{document}